\documentstyle[charm2000]{article}
\begin{document}


\title{
\begin{flushright}
HUTP-94/E021 \\
July 1994  \\
\end{flushright}
The $D^0\bar{D}^0$ Mixing Search --- Current
Status and Future Prospects~\thanks{To appear in the
Proceedings of the Charm 2000 Workshop, Fermilab, June 7-9, 1994.}}

\author{ Tiehui (Ted) Liu 		
      \\ {\sl High Energy Physics Laboratory}
       \\ {Harvard University, 42 Oxford Street}
       \\ {Cambridge, MA 02138}	
         }  
%

\maketitle

\vspace {-0.5 in}

\begin{abstract}

The search for $D^0\bar{D}^0$
mixing carries a large discovery potential for new
physics since the $D^0\bar{D}^0$
mixing rate is expected to be very small in the Standard
Model.  The past decade has seen significant experimental progress in
sensitivity, from 20\% down to 0.37\%.
This paper discusses the techniques,
current experimental status, and future prospects for the mixing
search. Some new ideas, applicable to future mixing searches,
are introduced.
The conclusion is that
while it is possible that the mixing sensitivity may
decrease to $10^{-5}$ around the year 2000, reaching the
$10^{-6}$ level will certainly be quite difficult.

\end{abstract}

\section{Introduction}
Following the discovery of the $D^0$ meson at SPEAR in 1976,
experimenters began to search for $D^0\bar{D}^0$ mixing, using
a variety of techniques. The past decade has seen
significant experimental progress in sensitivity
(from 20\%  to 0.37\%~\cite{SPEAR1}
to~\cite{CLEO15}), as can be seen in
Figure 1. Much of the enthusiasm for searching for $D^0\bar{D}^0$
mixing stems from the belief that the search carries a large discovery
potential for New Physics, since the mixing rate
${\rm R_{mixing}} \equiv {\cal B}(D^0 \to \bar{D}^0 \to \bar{f})$/
${\cal B}(D^0 \to f)$ is expected to be very small in the Standard Model.
One can characterize $D^0\bar{D}^0$ mixing in terms of two
dimensionless variables:
$ x={\delta m / \gamma_+}$ and $y={\gamma_- / \gamma_+}$, where
the quantities $\gamma_\pm$ and $ \delta m$ are defined by
$\gamma_\pm = {(\gamma_1\pm \gamma_2) / 2}$ and $ \delta m = m_2 - m_1$
with $m_i,\gamma_i$ $(i=1,2)$ being the masses and decay rates of
the two CP (even and odd) eigenstates.
Assuming a small mixing, namely,
$\delta m, \gamma_- \ll \gamma_+$ or $x,y\ll 1$, we have
${\rm R}_{\rm mixing}= (x^2 + y^2)/2 $.
Mixing can be caused either by $x \neq 0$ (meaning that mixing
is genuinely caused by the $D^0 - \bar{D}^0$ transition) or
by $y \neq 0$ (meaning mixing is caused by the fact that
the fast decaying component quickly disapears, leaving the slow
decaying component which is a mixture of $D^0$ and $\bar{D}^0$).
Theoretical calculations of $D^0\bar{D}^0$ mixing
in the Standard Model are plagued by large uncertainties. While
short distance effects from box diagrams are known~\cite{Gaillard}
to give a negligible contribution ($\sim 10^{-10}$),
the long distance effects from
second-order weak interactions with mesonic intermediate states
may give a much larger contribution. Estimates of
${\rm R}_{\rm mixing}$ from long distance effects range from $10^{-7}$
to $10^{-3}$~\cite{Donoghue}.
However, it has recently been argued by
Georgi and others that the long
distance contributions are smaller than previously estimated, implying
that cancellations occur between contributions from different
classes of intermediate mesonic states~\cite{Georgi}, and the
prevailing conclusion within the Standard Model seems to be that
${\rm R}_{\rm mixing}<10^{-7}$~\cite{Burdman}.
A measurement of such a small mixing rate is not possible with present
experimental sensitivity. However, the observation of a larger value
for ${\rm R}_{\rm mixing}$ caused by $x \neq 0$ would imply
the existence of new physics beyond the Standard Model~\cite{newphysics}.
Examples includes flavor-changing neutral currents mediated by the
exchange of a non-standard Higgs scalar with a mass of a few TeV/c$^2$,
which could lead to ${\rm R}_{\rm mixing}$ as large as $0.5\%$.

Recently, CLEO has observed a signal for $D^0\to K^+\pi^-$~\cite{footnote},
and found ${\rm R}$ = ${\cal B}(D^0 \to K^+\pi^-)/$
${\cal B}(D^0 \to K^-\pi^+) \sim 0.8\%$~\cite{Liu}.
Normally, $D^0$ decays by
Cabibbo favored decay $D^0 \to K^-\pi^+$ and $\bar{D}^0 \to K^+\pi^-$.
A signal for $D^0 \to K^+\pi^-$ could indicate mixing of $D^0 \to \bar{D}^0$.
But it could also indicate a different decay channel, namely,
Doubly Cabibbo Suppressed Decay(DCSD) $D^0 \to K^+\pi^-$,
which is suppressed with
respect to the Cabibbo favored decay
by a factor of $tan^4\theta_C \sim 0.3\%$ where
$\theta_C$ is the Cabibbo angle. Unfortunately,
without a precision vertex detector, CLEO is unable to
distinguish a potential mixing signal from DCSD.
As Purohit pointed out~\cite{Purohit}:
``The CLEO II result makes the entire subject of
$D^0\bar{D}^0$ mixing very interesting. It
really calls for a fixed-target experiment to use its decay time resolution to
decide whether the signal is due to DCSD or mixing". If the number of
reconstructed charm decays can reach $10^8$ around the year 2000, that would
allow one to reach a new threshold of sensitivity to
$D^0\bar{D}^0$ mixing, and perhaps actually observe it.
This is why charm mixing has been singled out for its own
working group at this workshop.

This paper is organized as follows: Section 2 discusses
the techniques which can be used to search for mixing, including two
new ideas. One of them is to use $D^0 \to K^+K^-,\pi^+\pi^-$ etc. to
study mixing (see 2.1.1), and the other is to use the difference in the
resonant substruture in $D^0 \to K^+\pi^-\pi^0$,$D^0 \to K^+\pi^-\pi^+\pi^-$
etc. to distingusih mixing and DCSD (see 2.1.2). In each case,
the relevant phenomenology will be briefly presented.
Section 3 discusses the present status and future prospects of
searching for mixing at different experiments. In section 4,
a comparison of the future prospects of the
different experiments with different
techniques will be given. A brief summary is in Section 5.

\section{The Techniques}
The techniques which can be used to search for mixing can be roughly
divided into two classes: hadronic and semi-leptonic. Each method has
advantages and limitations, which are described below.

\subsection{Hadronic method}
The hadronic method is to search for the $D^0$ decays
$D^0\to K^+\pi^-(X)$. These decays can occur either through
$D^0\bar{D}^0$ mixing followed by Cabibbo favored
decay $D^0 \to \bar{D}^0  \to K^+\pi^-$(X), or
through DCSD
$D^0\to K^+\pi^-(X)$. This means that the major complication for this
method is the need to distinguish between DCSD and mixing~\cite{early}.
The hadronic method can therefore be classified according to how
DCSD and mixing are distinguished. In principle, there are
at least three different ways to distinguish between DCSD and
mixing candidates experimentally:
(A) use the difference in the decay time-dependence;
(B) use the possible difference in the resonant substructure
    in $D^0 \to K^+\pi^-\pi^0$,$ K^+\pi^-\pi^+\pi^-$, etc. modes;
(C) use the quantum statistics of the production
and the decay processes.

Method (A) requires that the $D^0$ be highly boosted and so that
the decay time information can be measured. Method (B) requires
knowledge of the resonant
substructure of the DCSD decays, which is unfortunately something
about which we have no idea at this time.
Finally, method (C) requires that one use $e^+e^-$
annihilation in the charm threshold region.
In the following, we will discuss these three methods in some detail.

\subsubsection{Method A --use the
difference in the time-dependence of the decay}

This method~\cite{Bjorken} is to measure the decay time of the
$D^0 \to K^+\pi^-$ decay.
Here the $D^0$ tagging is usually done by using
the decay chain $D^{*+} \to D^0{\pi}_{\rm s}^+$ followed by
$D^0 \to K^+ \pi^-$. The ${\pi}_{\rm s}^+$ from $D^{*+}$ has a soft momentum
spectrum and is refered to as the slow pion.
The idea is to search for the wrong sign $D^{*+}$ decays, where the
slow pion has the same charge as the kaon arising from the $D^0$ decay.
This technique utilizes the following facts:
(1) DCSD and mixing have different decay time-dependence, which will
be described below.
(2) The charge of the
slow pion is correlated with the charm quantum number
of the $D^0$ meson and thus can be used to tag whether a $D^0$ or
$\bar{D}^0$ meson was produced in the decay
$D^{*+} \to D^0{\pi}_{\rm s}^+$ or $D^{*-} \to \bar{D}^0{\pi}_{\rm s}^-$.
(3) The small $Q$ value of the $D^{*+}$ decay
results in a very good mass resolution in the mass difference
$\Delta M \equiv M(D^{*+}) - M(D^0) - M({\pi}_{\rm s}^+)$ and allows
a $D^{*+}$ signal to obtained with very low background.
(4) The right sign signal $D^{*+} \to D^0{\pi}_{\rm s}^+$
followed by $D^0 \to K^- \pi^+$ can be used to provide
a model-independent normalization for the mixing measurement.

A pure $D^0$ state generated at $t=0$  decays to the $K^+\pi^-$ state either
by $D^0\bar{D}^0$ mixing or by DCSD, and the
two amplitudes may interfere.
The amplitude for a $D^0$
decays to $K^+\pi^-$ relative to the
amplitude for a $D^0$ decays to $K^-\pi^+$ is given by
\begin{equation}
\label{mixingamp}
A= \sqrt{{\rm R}_{\rm mixing}/2}\:\; t +
\sqrt{{\rm R}_{\rm DCSD}}\:\; e^{i\phi}
\end{equation}
where $\phi$ is an unknown phase, t is
measured in units of average $D^0$
lifetime.
Here ${\rm R}_{DCSD}=|\rho|^2$
where $\rho$ is defined as
$ \rho = {Amp(D^0\to K^+\pi^-)/Amp(\bar{D}^0\to K^+\pi^-) }$, denoting
the relative strength of DCSD.
We have also assumed a small mixing; namely,
$\delta m, \gamma_- \ll \gamma_+$ or $x,y\ll 1$, and
CP conservation.

The first term, which is proportional to $t$, is due to mixing
and the second term is due to DCSD. It is this unique attribute of the
decay time-dependence of mixing which can be used to distinguish
between DCSD and mixing. Now we have:
\begin{equation}
\label{mixing}
{\rm I}(D^0 \to K^+\pi^-)(t) \propto ({\rm R}_{\rm DCSD}
+\sqrt{2{\rm R}_{\rm mixing}{\rm R}_{\rm DCSD}}\;\: t cos\phi
+{1/2}\;{\rm R}_{\rm mixing}t^2)e^{-t}
\end{equation}
Define $\alpha = {\rm R}_{\rm mixing}/{\rm R}_{\rm DCSD}$, which
describes the strength of mixing relative to DCSD. Equation (2)
can then be rewritten as:
\begin{equation}
\label{mixing1}
{\rm I}(D^0 \to K^+\pi^-)(t) \propto {\rm R}_{\rm DCSD}(1+\sqrt{2\alpha}
\; t cos\phi +{1/2}\alpha t^2)e^{-t}
\end{equation}

{}From this equation, one may read off the following properties:
(1) The mixing term peaks at $t=2$.
(2) The interference term peaks at $t=1$.
(3) A small mixing signature can be greatly enhanced by DCSD
through interference (with $cos\phi \neq 0$) at lower
decay times. The ratio between the interference term and the mixing
term, denoted $\xi (t)$, is given by
$\xi (t)=\sqrt{8/\alpha}$ $cos\phi/t \propto \sqrt{1/\alpha}$. So
when $\alpha \rightarrow 0$, $\xi \rightarrow \infty$.
For example,  with $cos\phi$=1, at $t=1$
for $\alpha = 10\%, 1\%, 0.1\%, 0.01\%, 0.001\%$ (corresponding to
${\rm R}_{\rm mixing} = 10^{-3}, 10^{-4}, 10^{-5},10^{-6},10^{-7}$)
one has $\xi(t) = 9, 28, 90, 280,900$ respectively.
(4) Only for $t>\sqrt{8/\alpha}|cos\phi|$ does the interference
term become smaller than the mixing term.
(5) ${\rm I}(t_{0})=0$ happens and only happens when $cos\phi=-1$,
and only at location $t_{0}=\sqrt{2/\alpha}$.
For $\alpha = 10\%, 1\%, 0.1\%$, one has
$t_{0} = 4.5, 14.1, 44.7$.
(6) One can obtain a very pure DCSD sample by cutting at low decay time:
$t < \zeta \sim 0.1$. At such low $t$, the mixing term drops out and
leaving only the interference term. Let's define the purity of DCSD
to be ${\cal P}_{DCSD} = 1 - \int_{0}^{\zeta} {\sqrt{2\alpha}
\; cos\phi(te^{-t})}/\int_{0}^{\zeta} {e^{-t}}$. For
$\zeta \sim 0.1$ one get
${\cal P}_{DCSD} = \sqrt{2\alpha}cos\phi \:\zeta/(2+\zeta)$.
Let's take $\zeta=0.2$, for $\alpha = 10\%, 1\%$,
we can get ${\cal P}_{DCSD} = 96\%, 99\%$ pure DCSD respectively.
(7) The cut $t \leq \zeta$ cuts off only
$1-(1+\zeta)e^{-\zeta} \sim {\zeta}^2/2 = 2\%$ from the
whole interference term.

While Property (1) tells us that the mixing term does live at longer
decay time,
Property (3) tells us clearly that we should not ignore the interference
term. In fact, that's the last thing one wants to ignore! (unless
we know for sure $\cos\phi=0$).
It is the commonly believed ``annoying background'', namely DCSD, that
could greatly enhance the chance of seeing a very small mixing signal.
In semi-leptonic method, one does not have this advantage.
For a very small mixing rate, almost
all the mixing signature could show up in the interference term, not
in the mixing term, as long as $\cos\phi \neq 0$.
Property (2) tells us at which location one expect to find
the richest signature of a potential small mixing, which is
where the interference term peaks: $t \sim 1$
(why should one keep worrying about long lived DCSD tails?
let's hope for $\cos\phi \neq 0$ first.)
Property (5) shows that destructive interference is not necessarily
a bad thing.
In fact, it could provide extra information. For example, if
$\cos\phi = -1$, then one should find ${\rm I}(t_{0})=0$
at $t_{0}=\sqrt{2/\alpha}$, see Figure 5.
For the general case, interference will lead to very characteristic time
distribution, as can be clearly seen in Figure 6.
Properties (6) and (7) show that we can study DCSD well without being confused
by the
possible mixing component. This will become important when we discuss
method B.

Therefore the signature of mixing is
a deviation from a perfect exponential time distribution with the
slope of $\gamma_+$~\footnote{One can use $D^0 \to K^-\pi^+$
to study the acceptance function versus decay time.}.
Our ability to observe this signature depends on
the number of $D^0 \to K^+\pi^-$ events we will have.
Right now this is limited by the rather poor statistics.
Figures 3 and 4 show each term with $\alpha = 10\%$ and $cos\phi = \pm 1$
(with ${\rm R}_{\rm DCSD}=1$).

It is interesting to point out here that there is also a possibility,
previously unrecognized, of using the Singly Cabibbo
Suppressed Decays (SCSD), such as $D^0 \to K^+K^-,\pi^+\pi^-$
to study mixing. This is because (assuming CP conservation)
those decays occur only through the CP even eigenstate,
which means the decay time distribution is a perfect
exponential with the slope of $\gamma_1$.
Therefore, one can use those modes to measure $\gamma_1$.
The mixing signature is not a deviation from
a perfect exponential (again assuming CP conservation),
but rather a deviation of the
slope from $(\gamma_1 + \gamma_2)/2$. Since
${\gamma_+}= (\gamma_1 + \gamma_2)/2$ can be measured by
using the $D^0 \to K^-\pi^+$ decay time distribution, one can then
derive
$y={\gamma_- / \gamma_+}=(\gamma_1 - \gamma_2) /(\gamma_1 + \gamma_2)$.
Observation of a non-zero $y$ would demonstrate mixing caused
by the decay rate difference
(${\rm R}_{\rm mixing}= (x^2 + y^2)/2 $). It is worth pointing out
that in this case other CP even (odd) final states
such as $D^0 \to K^+K^-K^+K^-$ $(K^+K^-\pi^0)$
can be also used to measure $\gamma_1 (\gamma_2)$.
In addition, there is no need to tag the $D^0$ nor
know the primary vertex location, since we only need to determine
the slope. I should point out also that this method is only sensitive to
mixing caused by the decay rate difference between the two eigen states,
not to mixing caused by the mass difference $x={\delta m / \gamma_+}$
($\delta m = m_2 - m_1$).
The sensitivity of this method is discussed in section 4.1.

\subsubsection{Method B -- use difference in resonance substructure}

The idea of this new method is to use the wrong sign
decay $D^{*+} \to D^0{\pi}_{\rm s}^+$ followed by
$D^0 \to K^+ \pi^-\pi^0$, $K^+\pi^-\pi^+\pi^-$, etc., and use the
possible differences of the resonant substructure
between mixing and DCSD to study mixing.
There are good reasons to believe that the resonant substructure
of DCSD decay is different from that of mixing (Cabibbo favored
decay, CFD).
We can use the $D^0 \to K^+ \pi^-\pi^0$ decay as an example.
For CFD and DCSD,
the true yield density $n(p)$ at a point $p$ in the Dalitz plot
can be written as:
\begin{equation}
\label{density}
n(p) \:\propto \: \vert f_{1}\:e^{i\phi_{1}}A_{3b} +
f_{2}\:e^{i\phi_{2}}BW_{\rho^{+}}(p) + f_{3}\:e^{i\phi_{3}}BW_{K^{*-}}(p)
+f_{4}\:e^{i\phi_{4}}BW_{\bar{K}^{*0}}(p) {\vert}^2
\end{equation}
where $f_{i}$ are the relative amplitudes for each component and $\phi_{i}$
are the interference phases between each submode. $A_{3b}$ is
the S-wave three-body
decay amplitude, which is flat across the Dalitz plot. The various
$BW(p)$ terms are Breit-Wigner amplitudes for the $K^{*}\pi$ and
$K\rho$ sub-reactions.
Note that in general~\footnote{
It has been pointed out~\cite{Bigi,Bigi3} that it is unlikely that just one
universal suppression factor will affect the individual DCSD. For example,
SU(3) breaking can introduce a significant enhancement
for $D^0$ DCSD decays $D^0 \to K^+\pi^-, K^{*+}\pi^-$, while
SU(6) breaking can introduce a sizeable suppression relative to
the naive expectation for $D^0$ DCSD decay $D^0 \to K^+\rho^-$.},
\begin{equation}
\label{dcsd}
{f_{i}}^{DCSD}/{f_{i}}^{CFD}\neq {f_{j}}^{DCSD}/{f_{j}}^{CFD} \; (i \neq j)
\end{equation}
\begin{equation}
\label{phase}
{\phi_{i}}^{DCSD} \neq {\phi_{i}}^{CFD}
\end{equation}
This means that the resonant
substructure (the true yield density $n(p)$) for DCSD is different
from that of
mixing~\footnote{The sign of the interference between each submode
changes whenever $cos\theta_{R}$ ($\theta_{R}$ is the helicity angle
of the resonance) changes sign.
This is the same for both the Cabibbo favored decay and the
DCSD. This can be easily seen from the Breit-Wigner amplitude which
describes the strong resonances and decay angular momentum conservation:
$BW_{R} \:\propto \: \frac{cos\theta_{R}}{M_{ij}-M_{R}-i\Gamma_{R}/2}$ where
$M_{R}$ and $\Gamma_{R}$ are the mass and width of the $M_{ij}$
resonance (the $K^{*}$ or $\rho$).
The difference between Cabibbo favored decay and DCSD comes from
the relative amplitude $f_{i}$ for each submode and the interference
phase term $e^{i\phi_{i}}$.}.
As both DCSD and mixing contribute to the wrong sign
decay, the yield density for the wrong sign events
$n_w(p)$ will have a complicated form,
due to the fact that
for each submode DCSD and mixing may interfere with each other.
Just like in method A, for very small mixing, the interference
term between DCSD and mixing could be the most important one.

In principle, one can use the difference between the
resonant substructure for DCSD and mixing events to
distinguish mixing from DCSD. For instance,
combined with method A, one can perform a multi-dimensional
fit to the data by using the information on
$\Delta M$, $M(D^0)$, proper decay time $t$ and
the yield density on Dalitz plot
$n_{w}(p,t)$. The extra information on the
resonant substructure will, in principle, put a much better
constraint on the amount of mixing. Of course,
precise knowledge of the resonant substructure for DCSD
is needed here and so far we do not know anything
about it.
Because of this, for current experiments
this method is more likely to be a complication
rather than a better method
when one tries to apply method A to
$D^0 \to K^+ \pi^-\pi^0$ (see section 3.2 and~\cite{Liudpf})
or $D^0 \to K^+ \pi^-\pi^+\pi^-$.
In principle, however, one can use wrong sign samples at
low decay time (which is almost pure DCSD, see section 2.1.1.) to
study the resonant substructure of the DCSD decays.
In the near future, we should have a good understanding
of DCSD decays and this method could become a feasible
way to search for mixing.

\subsubsection{Method C ---use quantum statistics of the production
and decay processes}
This method is to search for dual identical two-body hadronic
decays in $e^+e^- \to \Psi'' \to D^0\bar{D}^0$, such as
$(K^-\pi^+)(K^-\pi^+)$, as was first suggested by Yamamoto
in his Ph.D thesis~\cite{Yamamoto}.
The idea is that when $D^0\bar{D}^0$ pairs are generated in a state
of odd orbital angular momentum (such as $\Psi''$), the DCSD
contribution to identical two-body pseudo-scalar-vector ($D \to PV$)
and pseudo-scalar-pseudo-scalar ($D \to PP$) hadronic decays
(such as $(K^-\pi^+)(K^-\pi^+)$) cancels out, leaving only the contribution
of mixing~\cite{Yamamoto,Bigi,Du}. As many people have asked about this,
I would like to show here the essence of Yamamoto's
original calculation for the
$(K^-\pi^+)(K^-\pi^+)$ case.
Let's define $e_i(t)=e^{-im_it- \gamma_i t/2}$ ($i=1,2$) and
$e_{\pm}(t) = (e_1(t) \pm e_2(t))/2$. A state that is
purely $|D^0 \rangle $ or $| \bar{D}^0 \rangle$ at time $t=0$ will evolve to
$|D(t) \rangle$ or $| \bar{D}(t)\rangle$ at time $t$, with
$|D(t) \rangle=e_+(t) |D^0 \rangle + e_-(t) |\bar{D}^0 \rangle$
and $|\bar{D}(t) \rangle = e_-(t) |D^0\rangle + e_+(t) |\bar{D}^0\rangle$.
In $e^+e^- \to \Psi'' \to D^0\bar{D}^0$,
the $D^0\bar{D}^0$ pair is generated in the state
$D^0\bar{D}^0 - \bar{D}^0D^0$ as the
relative orbital angular momentum of the pair
${\cal L} = 1$. Therefore, the time evolution of this state
is given by $|D(t)\bar{D}(t')\rangle - | \bar{D}(t)D(t') \rangle$,
where $t$ ($t'$) is the time of decay of the $D$ ($\bar{D}$).
Now the double-time amplitude ${\cal A}_w (t,t')$
that the left side decays to $K^-\pi^+$ at $t$ and
the right side decays to $K^-\pi^+$ at $t'$, giving
a wrong sign event $(K^-\pi^+)(K^-\pi^+)$, is given by:
\begin{equation}
\label{ddbarampw}
{\cal A}_w(t,t') = (e_+(t)e_-(t') - e_-(t)e_+(t'))(a^2 - b^2)
\end{equation}
where $a=\langle K^-\pi^+ | D^0 \rangle$ is the amplitude of the Cabibbo
favored decay $D^0 \to K^-\pi^+$, while $b=\langle K^-\pi^+|\bar{D}^0 \rangle$
is the amplitude of DCSD
$\bar{D}^0 \to K^-\pi^+$. Similarly, the double-time
amplitude ${\cal A}_r(t,t')$ for the right sign event
$(K^-\pi^+)(K^+\pi^-)$ is given by:
\begin{equation}
\label{ddbarampr}
{\cal A}_r(t,t') = (e_+(t)e_+(t') - e_-(t)e_-(t'))(a^2 - b^2)
\end{equation}
One measures the wrong sign versus right sign ratio ${\rm R}$, which is:
\begin{equation}
\label{nodcsd}
{\rm R}=\frac{{\rm N}(K^-\pi^+,K^-\pi^+)+{\rm N}(K^+\pi^-,K^+\pi^-)
}{{\rm N}(K^-\pi^+,K^+\pi^-)+{\rm N}(K^+\pi^-,K^-\pi^+) }
=\frac{\int\!\!\int {\cal A}_w(t,t')\,dt\,dt'}
{\int\!\!\int {\cal A}_r(t,t')\,dt\,dt'}
\end{equation}
Note in taking the ratio, the amplitude term $(a^2 - b^2)$ in Equations
(7) and (8) drop out. Thus, clearly $\rm R$ does not depend on
whether $b$ is zero (no DCSD) or finite (with DCSD).
Integrating over all times, one then obtains
${\rm R}=(x^2 + y^2)/2 = {\rm R}_{\rm mixing}$, where $x$ and $y$
are defined as before.

This is probably
the best way to separate DCSD and mixing.
The exclusive nature of the production guarantees both
low combinatoric backgrounds and production kinematics
essential for background rejection.
This method requires one use $e^+e^-$ annihilation
in the charm threshold region. Here the best final state is
$(K^-\pi^+)(K^-\pi^+)$.
In principle, one can also use final states like
$(K^-\rho^+)(K^-\rho^+)$ or $(K^{*-}\pi^+)(K^{*-}\pi^+)$,
etc., although again there are complications. For example, it is
hard to differentiate experimentally $(K^-\rho^+)(K^-\rho^+)$ from
$(K^-\rho^+)(K^-\pi^+\pi^0)$, where DCSD can contribute.
With high statistics, in principle, this method could
be combined with method B.

It has been pointed out that quantum statistics yield
different correlations for the $D^0\bar{D}^0$ decays from
$e^+e^- \to D^0\bar{D}^0, D^0\bar{D}^0\gamma,D^0\bar{D}^0\pi^0$~\cite{Bigi2}.
The well-defined coherent quantum states of the $D^0\bar{D}^0$
can be, in principle, used to provide valuable cross checks on systematic
uncertainties and, more importantly, to
extract $ x={\delta m / \gamma_+}$ and $ y={\gamma_- / \gamma_+}$
(which requires running at different energies)
if mixing is observed~\cite{Bigi2}.

\subsection{Semi-leptonic method}
The semi-leptonic method is to search for
$D^0\to \bar{D^0}\to X l^- \nu$ decays,
where there is no DCSD involved. However, it usually (not always!)
suffers from a large background due to the missing neutrino,
in addition, the need to
understand the large background often introduces model dependence.
In the early days, the small size of fully
reconstructed samples of exclusive $D^0$ hadronic decays and the lack of
the decay time information made it difficult to constrain the
$D^0\bar{D}^0$ mixing rate using the hadronic
method, many experiments used semi-leptonic decays.
The techniques that were used were similar
---searching for like-sign
$\mu^+\mu^+$ or $\mu^-\mu^-$ pairs in
$\mu^+N\to \mu^+(\mu^+\mu^+)X$~\cite{EMC,BDMS} and $\pi^- Fe \to
\mu^+\mu^+$~\cite{CCFRS}, $\pi^- W \to \mu^+\mu^+$~\cite{E615}.
These techniques rely
on the assumptions on production mechanisms, and the accuracy
of Monte Carlo simulations to determine the large conventional sources
of background.

There are other ways of using the semi-leptonic method.
The best place to use the semi-leptonic method is probably in
$e^+e^-$ annihilation near the charm threshold region.
The idea is to search for $e^+e^- \to \Psi'' \to $
$D^0\bar{D}^0 \to (K^-l^+\nu)(K^-l^+\nu)$ or
$e^+e^- \to D^-D^{*+} \to (K^+\pi^-\pi^-)(K^+\l^-\nu){\pi}_{\rm s}^+$
{}~\cite{Gladding3,Schindler}.
The latter is probably the only place where the semi-leptonic method does
not suffer from a large background. It should have a low background,
as there is only one neutrino missing in the entire event,
threshold kinematics constraints should provide clean signal.

It has been pointed out that one can not claim a
$D^0\bar{D}^0$ mixing signal based on the semi-leptonic method
alone~\footnote{
Bigi~\cite{Bigi2} has pointed out that an observation of
a signal on $D^0 \to l^-X$ establishes only that a certain selection
rule is violated in processes where the charm quantum number is changed,
namely the rule $\Delta {\rm {Charm}} = - \Delta {\rm Q_l}$ where
${\rm Q_l}$ denotes leptonic charge. This violation can occur
either through $D^0\bar{D}^0$ mixing (with the unique attribute of
the decay time-dependence of mixing)
, or through new physics beyond
the Standard Model (which could be independent of time).}
(unless with the information on decay time of $D^0$).
Nevertheless, one can always use this method to set upper limit
for mixing.

\section{Mixing Searches at Different Experiments}

\subsection{$e^+e^-$ running on $\Psi''(3770)$ --MARK III, BES,
Tau-charm factory}

The MARK III collaboration was the first (though
hopefully not the last)
to use the $e^+e^- \to \Psi'' \to D^0\bar{D}^0$ technique. They
reported preliminary results for two ``wrong-sign''
$D^0$ decay events (unpublished)~\cite{Gladding1},
one was consistent with $K^-\rho^+$ vs. $K^-\rho^+$, while the other
one was consistent with $K^{*0}\pi^0$ vs. $K^{*0}\pi^0$. This was a
very interesting result at that time, and had a strong influence
on the subject. However, one cannot draw a firm conclusion
about the existence of $D^0\bar{D}^0$ mixing based on these events.
There are at least two reasons:
(1) The background study has to rely on Monte Carlo
simulation of the PID (particle identification -- Time-of-Flight).
As Gladding has pointed out: ``These results must be considered preliminary
because the calculation of the confidence level is sensitive to the
tails of PID distribution for the background''~\cite{Gladding2};
(2) Assuming that the Monte Carlo background study is correct, and that
the events are real, one still cannot claim the two events are due to mixing,
for example,
the non-resonant decays $D^0 \to K\pi\pi^0$ may contribute to one
side of the pair in each of the events, in which DCSD can contribute.

The MARK III puzzle can be completely solved at a $\tau$-charm factory,
which is a high luminosity ($10^{33} cm^{-2}s^{-1}$) $e^+e^-$ storage ring
operating at center-of-mass energies in the range 3-5 GeV.
The perspectives for a $D^0\bar{D}^0$ mixing search at a
$\tau$-charm factory have been studied in some
detail~\cite{Gladding3,Schindler}.
I will outline here the most important
parts. The best way to search for mixing is probably to use
$e^+e^- \to \Psi'' \to D^0\bar{D}^0 \to (K^-\pi^+)(K^-\pi^+)$.
The sensitivity is not hard to estimate. Assuming a one year run
with a luminosity of $10^{33} cm^{-2}s^{-1}$, $5.8 \times 10^7$ $D^0$s
would be produced from $\Psi''$. Therefore about $9 \times 10^4$
$(K^-\pi^+)(K^+\pi^-)$ events would be produced.
About $40\%$ of them ($3.6 \times 10^4$)
could be fully reconstructed.
A detailed study has shown that the potential dominant background
comes from doubly misidentified $(K^-\pi^+)(K^+\pi^-)$, and if
TOF resolution is 120 ps, this background can
be kept to the level of one event or less. This means one can set an
upper limit at the $10^{-4}$ level.

As mentioned section 2.2, the best place to use the semi-leptonic method
is probably at a $\tau$-charm factory.
One good example is to search for $e^+e^- \to D^-D^{*+} \to$
$(K^+\pi^-\pi^+)(K^+l^-\nu){\pi}_{\rm s}^+$. It is expected that
this method can also have a sensitivity at the $10^{-4}$ level.
There are many other independent techniques that one can use for a mixing
search at a $\tau$-charm factory. By combining several independent
techniques (which require running at different energies), it was claimed that
$D^0\bar{D}^0$ mixing at the $10^{-5}$ level could be
observable~\cite{Schindler}.

There have been several schemes around the world for building a
$\tau$-charm factory. If such a machine is built, it could be
a good place to study mixing. At the workshop, Walter Toki told
us the history of the $\tau$-charm factory: one was proposed at SLAC
in 1989 and at Spain in 1993. It was discussed at
Dubna in 1991, at IHEP (China), and at Argonne (this workshop).
It will be discussed again at IHEP (China) soon.
Let us hope that we will have one in the not too distant future.

\subsection{$e^+e^-$ running near $\Upsilon(4S)$ --ARGUS, CLEOII, CLEO III,
B factory}

Using the CLEO II data sample,
with an integrated luminosity of 1.8 fb$^{-1}$
at and near the $\Upsilon$(4S) resonance,
CLEO has observed a signal for
$D^0 \to K^+ \pi^-$~\cite{Liu} from the decay chain
$D^{*+} \to D^0{\pi}_{\rm s}^+ \to (K^+\pi^-){\pi}_{\rm s}^+$,
as can be seen in Figure 2.

Without a precision vertex detector, CLEO II can only in effect
measure the rate ${\cal B}(D^0\to K\pi)$
integrated over all times of a pure $D^0$ decaying to a final
state $K\pi$. The ratio
${\rm R}$=${\cal B}(D^0\to K^+ \pi^-)$ /${\cal B}(D^0 \to K^- \pi^+)$
is given by integrating equation (2) over all times:

\begin{equation}
\label{labeled-equation}
{\rm R}= {\rm R}_{\rm mixing} + {\rm R}_{\rm DCSD}
+\sqrt{2{\rm R}_{\rm mixing}{\rm R}_{\rm DCSD}} cos\phi
\end{equation}

CLEO II finds $ {\rm R}=
(0.77\pm 0.25\,({\rm stat.})\pm 0.25\,({\rm sys.}))\% $.
This signal could mean one of two things: (1) mixing could be quite large,
which would imply that mixing can be observed in the near future;
(2) the signal is dominated by DCSD.
The theoretical prediction for ${\rm R_{DCSD}}$ is about
2 $tan^4\theta_C \sim 0.6\%$~\cite{Bigi,Chau}, which is quite
consistent with the measured value. It is, therefore,
believed by many that the signal is due to DCSD,
although it remains consistent with
the current best experimental upper limits on mixing, which
are $(0.37-0.7)\%$~\cite{Browder} and $0.56\%$~\cite{E615}.

CLEO has also tried to use hadronic method B, by
searching for $D^0 \to K^+\pi^-\pi^0$.
The excellent photon detection at CLEO II allows one to study this mode with
a sensitivity close to $D^0 \to K^+\pi^-$ mode.
The main complication faced here is that (as discussed in section 2.1.2)
the resonant substructure is not
necessarily the same for wrong sign and right sign decays.
Because of this, the interpretation of ${\rm R}$ as ${\rm R_{mixing}}$
or ${\rm R_{DCSD}}$ will be complicated by the lack of knowledge
of the details of the interference between submodes (and also the
decay time information).
Moreover, one has to worry about the detection efficiency
across the Dalitz plot. Setting an upper limit for each
submode is clearly very difficult.
CLEO has thus set an upper limit on the inclusive rate
for $D^0 \to K^+\pi^-\pi^0$ as
${\rm R}$ = ${\cal B}(D^0 \to K^+\pi^-\pi^0)$
/${\cal B}(D^0 \to K^-\pi^+\pi^0)$ $< 0.68\%$~\cite{Liudpf}.
Note this upper limit includes the possible effects of the
interference between the DCSD and mixing for each submode
as well as the interference between submodes.

This summer, CLEO will install a silicon vertex detector (SVX)
with a longitudinal resolution on
vertex separation around 75 $\mu {\rm m}$. This will
enable CLEO to measure the
decay time of the $D^0$, and reduce the random slow pion background
(as the resolution of the $D^{*+}$ - $D^0$ mass difference is dominated
by the angular resolution on the slow pion, this should be
greatly improved by the use of the SVX).
By the year 2000, with CLEO III (a symmetric B factory)
and asymmetric B factories at SLAC and KEK,
each should have about thousands $D^0 \to K^+K^-(X),\pi^+\pi^-(X)$
and a few hundred $D^0 \to K^+\pi^-$
(and perhaps $D^0 \to K^+\pi^-\pi^0$, $K^+\pi^-\pi^+\pi^-$ too)
signal events with decay time information for one year of running.
The typical decay length of
$D^0$ (${\cal L}$) is about a few hundred $\mu {\rm m}$, and the resolution
of the decay length ($\sigma_{\cal L}$) is about 80 $\mu {\rm m}$
(${\cal L}/{\sigma_{\cal L}} \sim 3$). The sensitivity to mixing at
CLEO III and asymmetric B factory has not been carefully studied yet.
A reasonable guess is that it could be as low as $10^{-4}$.
If mixing is indeed as large as DCSD, it should be observed by then.

\subsection{Fixed target experiments--E615,E691, E791, E687,E831}
A significant amount of our knowledge has been gained from
Fermilab fixed target experiments, and in fact the current best upper
limits on mixing have emerged from these experiments (E615, E691), and
will come from their successors E687, E791 and E831 soon.

The best upper limit using the semi-leptonic method comes
from the Fermilab experiment E615, which used a 255 GeV pion beam on
a tungsten target. The technique is to search
for the reaction $\pi N \to D^0\bar{D}^0 \to$ $ (K^-\mu^+\nu)D^0$
$\to (K^-\mu^+\nu)(K^-\mu^+\nu)$, where only the final state
muons are detected (like sign $\mu^+\mu^+$ or $\mu^-\mu^-$ pairs).
Assuming $\sigma(c\bar{c}) \sim A^1$ nuclear dependence, they
obtained ${\rm R}_{\rm mixing} < 0.56\%$~\cite{E615}.

The best upper limit using the hadronic method by measuring the
decay time information comes from E691, which is the first high statistics
fixed target (photoproduction) experiment.
In fact, E691 was the first experiment which used the
decay time information (obtained from the excellent decay time
resolution of their silicon detectors) to distinguish DCSD and mixing.
The decay chains $D^{*+} \to D^0{\pi}_{\rm s}^+$ followed by
$D^0 \to K^+ \pi^-$, $K^+ \pi^-\pi^+\pi^-$ were used.
The upper limits from the $D^0 \to K^+\pi^-$ mode
are ${\rm R}_{\rm mixing} < (0.5-0.9)\%$ and ${\rm R}_{DCSD} < (1.5-4.9)\%$ ,
while the upper limits from $D^0 \to K^+\pi^-\pi^+\pi^-$
are ${\rm R}_{\rm mixing}<(0.4-0.7)\%$ and ${\rm R}_{DCSD} < (1.8-3.3)\%$ .
The combined result gives ${\rm R}_{\rm mixing} < (0.37-0.7)\%$.
The ranges above reflect the possible
effects of interference between DCSD and mixing with an unknown phase($\phi$).
Note that for $D^0 \to K^+ \pi^-\pi^+\pi^-$,
the resonant substructure in the Cabibbo favored and
DCSD decays has been ignored.

At this workshop, both E687 and E791 have reported their preliminary
result from part of their data. One can find the details in Jim Wiss's talk.
The best upper limits on mixing should come from these two
experiments soon.
It is worth pointing out here that both the E687 and E791 results
reported in the workshop are based on the assumption that there is no
interference between DCSD and mixing.
Future analysis should include the interference term for the reasons
discussed in section 2.1.1.

\section{Comparison of Different Experiments}

\subsection{Hadronic method A}

This measurement requires:
(1) excellent vertexing capabilities, at least good enough
to see the interference structure;
(2) low background around the primary vertex.
The background level around the primary vertex
could be an important issue as the interference term in equation (2)
does peak at $t=1$.
In addition, low background around primary vertex
means that one does not suffer much from random slow pion background
and also one can measure the DCSD component at low decay time well.
This is important for understanding DCSD at large decay times.
The vertexing capabilities at $e^+e^-$ experiments
(${\cal L}/\sigma \sim 3$) for CLEO III and asymmetric B factories
at SLAC and KEK should be sufficient for a mixing search.
The extra path-length due to the Lorentz boost,
together with the use of silicon detectors for high
resolution position measurements, have given the fixed
target experiments an advantage over $e^+e^-$ experiments
(${\cal L}/\sigma \sim 8-10$).
The low background around the primary vertex at
$e^+e^-$ experiments and photoproduction experiments
is a certain advantage.
It is worth pointing out here that at the
$e^+e^-$ experiments (esp. at asymmetric B factory or Z factory)
it maybe possible to use $B^0 \to D^{*+} l^-\nu $, where the
primary ($D^{*+}$ decay) vertex can be determined by the $l^-$ together with
the slow pion coming from the $D^{*+}$. In this case, the background level
around the primary vertex is intrinsically very low.

However, in the case of $D^0 \to K^+K^-,\pi^+\pi^-$, etc.,
the requirement on the background level around the primary
vertex is not so important. In this case,
the mixing signature is not a deviation from
a perfect exponential (again assuming CP conservation),
but rather a deviation of the
slope from $(\gamma_1 + \gamma_2)/2$.
It is worth pointing out that there are many advantages with
this method. For example, one can use Cabibbo favored decay modes,
such as $D^0 \to K^-\pi^+$, to measure the average $D^0$ decay rate
$(\gamma_1 + \gamma_2)/2$. This, along with other SCSD CP even (or odd)
final states, would allow for valuable cross checks on
systematics uncertainties. In addition, since we only need to determine
the slope here, we do not need to tag the $D^0$ nor
know the primary vertex location.
The sensitivity of this method
depends on how well we can dertermine the slope.
Roughly speaking, in the ideal case,
the sensitivity to $y$ would be $\sim 1/\sqrt{N}$, where $N$
is the number of $D^0 \to K^+K^-, \pi^+\pi^-$, etc. events,
which means that the sensitivity to mixing caused by
the decay rate difference ($\sim y^2/2$) would be close to
$\sim 1/N$.  For example,
a fixed-target experiment capable of
producing $\sim 10^8$ reconstructed charm events could lower
the sensitivity~\footnote
{Since I came up with this idea the day before the deadline of this paper,
the sensitivity has not been carefully estimated yet.}
to $\sim 10^{-5}-10^{-6}$ level for
the $y^2$ term in ${\rm R}_{\rm mixing}= (x^2 + y^2)/2$.

\subsection{Hadronic method B}

In the near future, we should be able to have a good understanding
of DCSD~\footnote{It may be possible that
good understanding of DCSD can be reached by measuring
the pattern of $D^+$ DCSD decays where the signature is not confused
by a mixing component. It is worth pointing out that the
$D^+$ DCSD decays can be studied very well at future
fixed target experiments.}
in $D^0 \to K^+\pi^-\pi^0$, $D^0 \to K^+\pi^-\pi^+\pi^-$, etc. modes,
then method B will
become a feasible way to study mixing and the sensitivity should be
improved. Just like method A, this method requires
very good vertexing capabilities and very low
background around the primary vertex (this is even more
important than in method A, since precise knowledge of DCSD is very
important here).
In addition, this method requires that the detection efficiency (for the
mode being searched) across Dalitz plot be quite
uniform (at least the detector should have good
acceptance on the Dalitz plot at locations where DCSD and mixing
resonant substructure are different). This is necessary so that
detailed information on the resonant substructure
can be obtained in every corner on the Dalitz plot.

The excellent photon detection capabilities will allow
$e^+e^-$ experiments to study the $D^0 \to K^+\pi^-\pi^0$ mode with very low
background. From the CLEO II $D^0 \to K^+\pi^-\pi^0$ analysis,
the detection efficiency across the Dalitz plot will have some
variations due to cuts needed to reduce background,
however, it is still good enough to obtain detailed information
on the resonant substructure~\cite{Liudpf}.
Future fixed target experiments may have a good chance to
study $D^0 \to K^+\pi^-\pi^+\pi^-$ mode, since the detection efficiency
across Dalitz plot should be quite flat.
The sensitivity that each experiment can reach by using this method
depends on many things and need to be carefully studied in the future.

\subsection{Hadronic method C}

The sensitivity of this method depends crucially on the
particle identification capabilities. Since the $D^0$ is at rest,
the $K$ and $\pi$ mesons will have the same momentum, so a
doubly misidentified Cabibbo favored decay $D^0 \to K^-\pi^+$
($K^- \to \pi^-, \pi^+ \to K^+$ ) mimics a $D^0 \to K^+\pi^-$
with almost the same $D^0$ mass. It is worth pointing out here
that particle identification is not as crucial to method A
as it is to this method (C), as far as this particular background is
concerned. This is because in method A, the $D^0$
is highly boosted, and doubly misidentified $D^0 \to K^+\pi^-$ decays
will have a broad distribution in the $D^0$ mass spectrum around the
$D^0$ mass peak; this background can be kinematically rejected
with only a small reduction of the efficiency for the signal
events~\footnote{The idea is for each $K^+\pi^-$ candidates,
one can invert the kaon and pion assignments and recalculate the $D^0$ mass,
denoted $M_{\rm flip}$. If $M_{\rm flip}$ lies close to the
nominal $D^0$ mass, the combination is discarded. This veto works
as long as the momentum measurement is correct. One can say that
excellent tracking capabilities is crucial in order to get rid of
this background here.}.

Once the sensitivity reaches ${\cal O}(10^{-5})$, one may have to
worry about other contributions, such as contributions from
continuum background, contributions from
$e^+e^- \to 2 \gamma \to D^0\bar{D^0}$ which may produce C-even
states where DCSD can contribute~\cite{Du}.

\subsection{Semi-leptonic method}

The semi-leptonic method usually suffers from large
background (except at a $\tau$ charm factory), the traditional
method of looking for like sign $\mu^+\mu^+$ or $\mu^-\mu^-$ pairs
is an example. New ideas are needed in order to improve the
sensitivity significantly. Some promising techniques
have been suggested by Rolly Morrison and
others, and have been discussed in the working group~\cite{Liu1}.

\section{Summary}

The search for $D^0\bar{D}^0$
mixing carries a large discovery potential for new
physics since the $D^0\bar{D}^0$
mixing rate is expected to be very small in the Standard
Model.  The past decade has seen significant experimental progress in
sensitivity (from 20\% down to 0.37\%).

In light of the recent CLEO II signal in $D^0 \to K^+\pi^-$,
if the mixing rate is close to that of DCSD (above $10^{-4}$)
, then it might be observed
by the year 2000 with either the hadronic or the semi-leptonic method,
either at fixed target experiments,
CLEO III, asymmetric B factories(at SLAC and KEK), or
at a $\tau$-charm factory.
If the mixing rate is indeed much smaller than DCSD,
then the hadronic method may have a better chance
as the
potentially very small mixing signature could be
enhanced by the presence of the relatively large
``annoying background'' DCSD.
The design of future experiments should focus on improving the
vertexing capabilities and reducing the
background level around the primary vertex, in order to
fully take advantage of having the possible DCSD interference.
In addition, the very complication due to
the possible differences between the
resonant substructure in many DCSD and mixing decay modes
$D^0 \to K^+\pi^-(X)$ could in principle be turned to advantage by
providing additional information once the substructure in DCSD
is understood (method B) and the sensitivity could be
improved significantly this way.
This means that understanding DCSD in $D$ decays could be a very important
step on the way to observe mixing. Experimenters
and theorists should work hard on this.

In the case of $D^0 \to K^+\pi^-(X)$ and $D^0 \to X^+ l^-$, we are only
measuring ${\rm R}_{\rm mixing}= (x^2 + y^2)/2 $.
Since many extensions of the Standard Model predict large
$ x={\delta m / \gamma_+}$, it is very important to measure
$x$ and $y$ separately. Fortunately, SCSD can provide us information
on $y$. This is due to the fact
that decays such as $D^0 \to K^+K^-,\pi^+\pi^-$,
occur only through definte CP eigenstate,
and this fact can be used to measure the decay rate difference
$y={\gamma_- / \gamma_+}=(\gamma_1 - \gamma_2) /(\gamma_1 + \gamma_2)$
alone. Observation of a non-zero $y$ would demonstrate mixing caused
by the decay rate difference.
This, together with the information on
${\rm R}_{\rm mixing}$ obtained from other methods, we can
in effect measure $x$. In this sense, it is best
to think of the quest to observe mixing (new physics) as a program rather than
a single effort.

\section*{Acknowledgement}

Much of my knowledge on this subject has been gained by
having worked closely with Hitoshi Yamamoto over the past three years,
I would like to express my sincere gratitude to him. Besides,
I would like to express special thanks to
Rolly Morrison for many useful and stimulating discussions,
to Richard Wilson for encouragement and careful reading of the draft,
to Gary Gladding, Tom Browder and Milind Purohit
for helping me to understand the details of
MARK III, E691 and E791 analysis (respectively), to Dan Kim
who made Figure 1.
Finally, I would like to thank the organizers of the workshop
for providing me the opportunity to write this paper.

\begin{figure}[htb]
\vspace{1.5in}
\begin{picture}(500,520)(0,0)
\end{picture}
\caption {The quest
for $D^0\bar{D}^0$ mixing. Note that the range in E691 result
reflects the possible effects of interference between DCSD and mixing,
and the CLEO II signal could be due to either mixing or DCSD,
or a combination of the two.}
\label{fig:history}
\end{figure}

\begin{figure}[htb]
\vspace{2.0in}
\begin{picture}(450,450)(0,0)
\end{picture}
\caption {The CLEO II signal for $D^0 \to K^+\pi^-$.
The $D^0$ mass for wrong sign events. (a) for events
in the $\Delta M$ peak; (b) for events in the
$\Delta M$ sidebands. The solid lines are the fits using the
corresponding right sign mean and $\sigma$ in data.}
\label{fig:signal}
\end{figure}

\begin{figure}[htb]
\vspace{2.0in}
\begin{picture}(450,450)(0,0)
\end{picture}
\caption {The decay time dependence of DCSD and mixing with
$\alpha = {\rm R}_{\rm mixing}/{\rm R}_{\rm DCSD}=10\%$.}
\label{fig:decaytime}
\end{figure}

\begin{figure}[htb]
\vspace{2.0in}
\begin{picture}(450,450)(0,0)
\end{picture}
\caption {The decay time dependence of DCSD and mixing
with $\alpha = {\rm R}_{\rm mixing}/{\rm R}_{\rm DCSD}=10\%$, in log scale.}
\label{fig:decaytimelog}
\end{figure}

\begin{figure}[htb]
\vspace{1.5in}
\begin{picture}(450,450)(0,0)
\end{picture}
\caption {The decay time dependence of DCSD and mixing
with maximal destructive interference $cos\phi=-1.0$.
For different $\alpha = {\rm R}_{\rm mixing}/{\rm R}_{\rm DCSD}$ values:
from left to right, $\alpha=100\%, 50\%, 10\%$,
$5\%, 1\%,0.5\%$ (with ${\rm R}_{\rm DCSD}=10^{-2}$,
this corresponds to ${\rm R}_{\rm mixing}=$ $10^{-2},
5 \times 10^{-3},10^{-3}$,
$5 \times 10^{-4},10^{-5},5 \times 10^{-6}$).}
\label{fig:alpha}
\end{figure}

\begin{figure}[htb]
\vspace{2.0in}
\begin{picture}(450,450)(0,0)
\end{picture}
\caption {The decay time dependence of DCSD and mixing
with $\alpha = {\rm R}_{\rm mixing}/{\rm R}_{\rm DCSD}=10\%$.
For different $cos\phi$ values: from bottom to top,
$cos\phi$ $=-1.0,-0.99,-0.96,-0.94,-0.80,$
$-0.6, 0.0, 0.5, 0.7, 1.0$. The solid line is the DCSD term,
as a reference line.}
\label{fig:cosphi}
\end{figure}


\begin{thebibliography}{99}
\bibitem{SPEAR1}
G.J.Feldman {\it et al.},{\it Phys.\ Rev.\ Lett.} {\bf 38}, 1313(1977).

\bibitem{SPEAR2}
G.Goldhaber {\it et al.}, {\it Phys.\ Lett.} B {\bf 69}, 503 (1977).

\bibitem{E87}
P.Avery {\it et al.},{\it Phys.\ Rev.\ Lett.} {\bf 44}, 1309(1980).

\bibitem{EMC}
J.J.Aubert {\it et al.}, {\it Phys.\ Lett.} B {\bf 106}, 419 (1981).

\bibitem{CCFRS}
A.Bodek {\it et al.},{\it Phys.\ Lett.} B {\bf 113}, 82 (1982).

\bibitem{ACCMOR}
R.Bailey {\it et al.}, {\it Phys.\ Lett.} B {\bf 132}, 237 (1983).

\bibitem{BDMS}
A.Benvenuti {\it et al.}, {\it Phys.\ Lett.} B {\bf 158}, 531 (1985).

\bibitem{DELCO}
H.Yamamoto {\it et al.}, {\it Phys.\ Rev.\ Lett.} {\bf 54}, 522 (1985).

\bibitem{HRS}
S.Abachi {\it et al.}, {\it Phys.\ Lett.} B {\bf 182}, 101 (1986).

\bibitem{E615}
W.C.Louis {\it et al.}, {\it Phys.\ Rev.\ Lett.} {\bf 56}, 1027 (1986).

\bibitem{ARGUS}
H. Albrecht {\it et al.}, {\it Phys.\ Lett.} {\bf 199}, 447 (1987).

\bibitem{Browder}J.C. Anjos {\it et al.}, (E691 Collaboration),
{\it Phys.\ Rev.\ Lett.} {\bf 60}, 1239 (1988). For details of this study,
see T. Browder, Ph.D. Thesis, UCSB-HEP-88-4, (1988).

\bibitem{CLEO15}R. Ammar {\it et al.}, (CLEO Collaboration), {\it Phys.\ Rev.}
 D {\bf 44}, 3383 (1991).


\bibitem{Gaillard}
M.K. Gaillard and B.W. Lee, {\it Phys.\ Rev.} D {\bf 10}, 897 (1974);
A. Datta, {\it Phys.\ Lett.} B {\bf 154}, 287 (1985).

\bibitem{Donoghue}
L. Wolfenstein, {\it Phys.\ Lett.} B {\bf 164}, 170 (1985);
J. Donoghue, E. Golowich, B.R. Holstein and J. Trampetic, {\it Phys.\ Rev.}
D {\bf 33}, 179 (1986).

\bibitem {Georgi}
H. Georgi, {\it Phys.\ Lett.} B {\bf 297}, 353 (1992);
T. Ohl, G. Ricciardi and E.H. Simmons, {\it Nucl.\ Phys.} B {\bf 403},
603 (1993).

\bibitem {Burdman}
G. Burdman, ``Charm mixing and CP violation in the Standard Model'',
S. Pakvasa, ``Charm as Probe of New Physics'', these Proceedings.

\bibitem {newphysics}
A. Datta,  {\it Phys.\ Lett.} B {\bf 154}, 287 (1985);
A.M. Hadeed and B. Holdom, {\it Phys.\ Lett.} B {\bf 159}, 379(1985);
B. Mukhopadhyaya, A. Raychaudhuri and A. Ray,
{\it Phys.\ Lett.} B {\bf 190}, 93(1987);
K.S. Babu {\it et al.}
{\it Phys.\ Lett.} B {\bf 205}, 540(1988);
E. Ma {\it et al.}
{\it Mod. \ Phys.\ Lett.}  A{\bf 3}, 319(1988);
L. Hall and S. Weinberg,  {\it et al.}
{\it Phys.\ Rev.} D {\bf 48}, 979 (1993).

\bibitem{footnote}We discuss $D^0$ decays explicitly in the text,
its charge conjugate decays are also implied
throughout the text unless otherwise stated.

\bibitem{Liu}D.Cinabro {\it et al.}, (CLEO Collaboration),
{\it Phys.\ Rev.\ Lett.} {\bf 72}, 1406 (1994).

\bibitem{Purohit}M. Purohit, ``A $D^0\bar{D}^0$ mixing and DCSD analysis
of E791 data'', E791 memo. March, 1994.

\bibitem{early}
In the early days when people searched for
$D^0\bar{D}^0$ mixing using the hadronic method,
the lack of knowledge on DCSD decays and
the lack of statistics and decay time information have forced one to
assume that DCSD is negligible
in order to set upper limit for
mixing~\cite{SPEAR1,SPEAR2,E87,ACCMOR,DELCO,HRS,ARGUS,CLEO15}.

\bibitem{Bjorken}
This idea was first suggested by J.D. Bjorken in 1985,
private communication with Tom Browder and Mike Witherell.

\bibitem{Bigi}I. Bigi and A.I. Sanda,
{\it Phys.\ Lett.} B {\bf 171}, 320 (1986).


\bibitem{Liudpf}
T.Liu, ``$D^0\bar{D}^0$ mixing and DCSD -
Search for $D^0 \to K^+\pi^-(\pi^0)$'',
to appear in the Proceedings of the Eighth Meeting of the Division
of Particles and Fields of the American Physics Society (DPF'94),
Albuquerque, New Mexico. (World Scientific).




\bibitem{Yamamoto} H. Yamamoto,  Ph.D. Thesis, CALT-68-1318 (1985).

\bibitem{Du} D. Du and D. Wu,
{\it Chinese Phys.\ Lett.} {\bf 3}, No. 9(1986).



\bibitem{Gladding3}
G.E. Gladding, in Proceedings of the $\tau$-charm factory worshop,
SLAC-Report-343, June 1989.

\bibitem{Schindler}
R.H. Schindler, in Proceedings of the $\tau$-charm factory worshop,
SLAC-Report-343, June 1989.




\bibitem{Liu1}
T.Liu, ``Charm mixing working group summary report'',
these Proceedings.

\bibitem{Gladding1}
G.E. Gladding, in Proceedings of the Fifth
International Conference on Physics in Collision, Autun,
France, 1985 (World Scientific).

\bibitem{Gladding2}
G.E. Gladding, ``$D^0\bar{D}^0$ Mixing, The Experimental Situation''
in Proceedings of
Internatioal Symposium on Production and Decay of Heavy Flavors,
Standford, 1988.


\bibitem{Bigi2}
I.Bigi, in Proceedings of the $\tau$-charm factory worshop,
SLAC-Report-343, June 1989.

\bibitem{Bigi3}I.Bigi, in SLAC summer institute on Particle
Physics, Stanford, 1987; edited by E.C. Brennan (SLAC, 1988);
Other predictions can be found in R.C. Verma and A.N. Kamal,
{\it Phys.Rev.} {\bf D43},829(1991).




\bibitem{Chau}L.L. Chau and
H.Y.Cheng, ITP-SB-93-49, UCD-93-31 (HEP-PH-9404207).






\end{thebibliography}
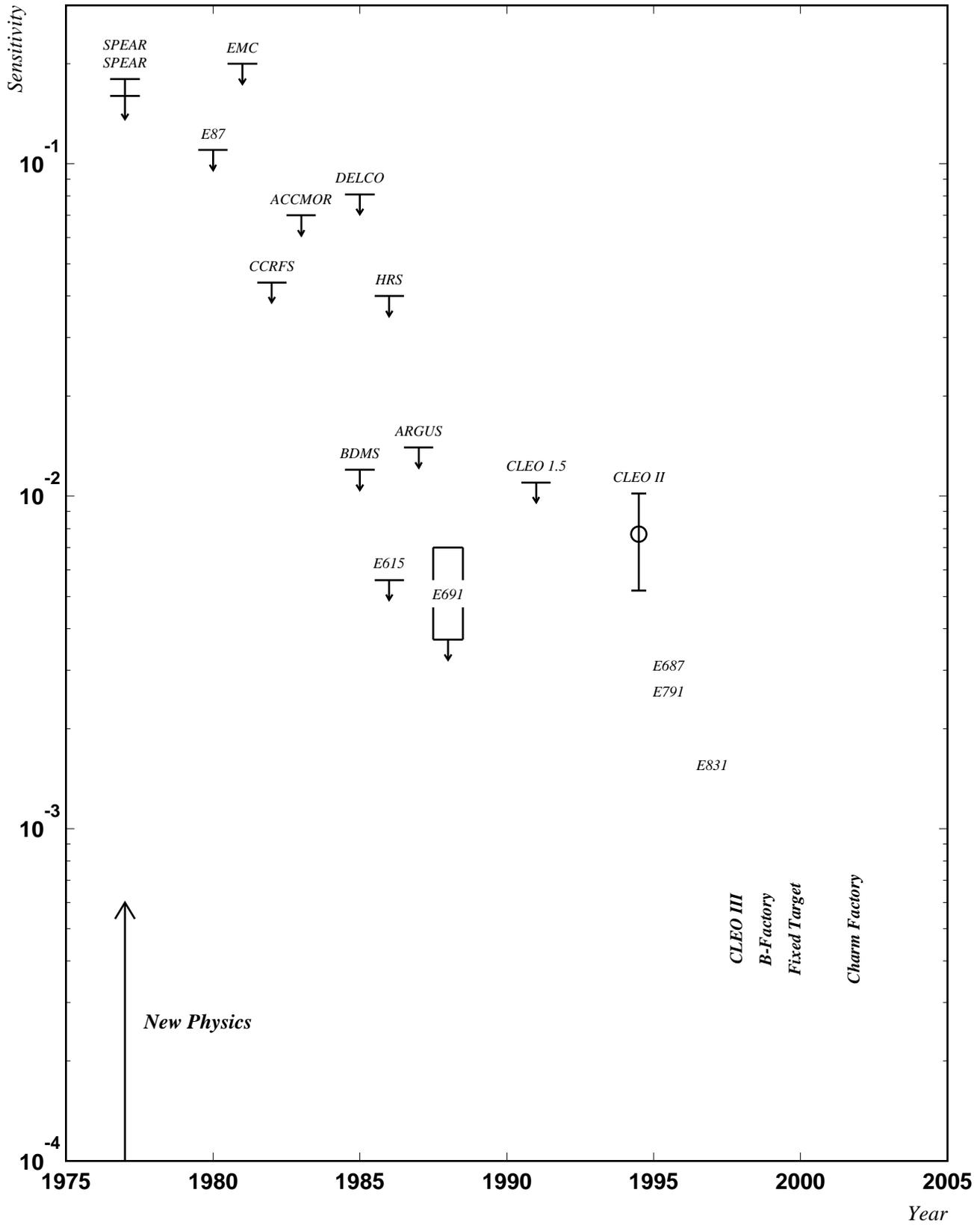
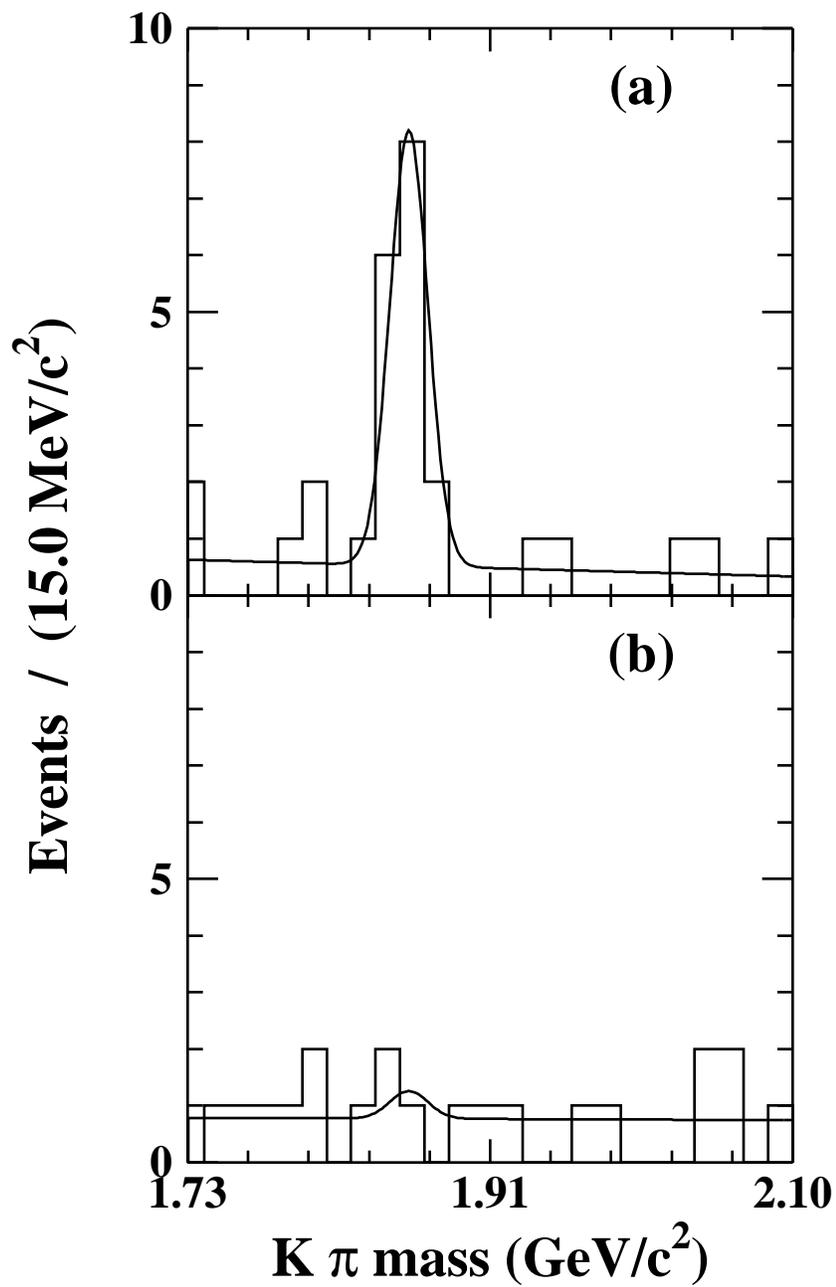
\end{document}